\documentclass[aps,preprint,prl,showpacs]{revtex4}

\usepackage{amsmath}
\usepackage{amssymb}
\usepackage{graphicx}

\begin{document}

\title{Intracycle and Intercycle Interferences in Above-Threshold
  Ionization: the Time Grating} \author{Diego G. Arb\'{o}$^{1,2}$,
  Kenichi L. Ishikawa$^{3}$, Klaus Schiessl$%
  ^{4}$, Emil Persson$^{4}$, and Joachim Burgd\"{o}rfer$^{4}$}
\affiliation{$^{1}$Institute for Astronomy and Space Physics, IAFE (Conicet-UBA), CC
  67, Suc. 28 (1428)
  Buenos Aires, Argentina\\
  $^{2}$Department of Physics, FCEN, University of Buenos Aires, Argentina\\
  $^{3}$Photon Science Center, Graduate School of Engineering,
  the University of
  Tokyo, Hongo 7-3-1, Bunkyo-ku, Tokyo 113-8656, Japan\\
  $^{4}$Institute for Theoretical Physics, Vienna University of
  Technology, Wiedner Hauptstra\ss e 8-10/136, A-1040 Vienna, Austria,
  EU} \date{\today}

\begin{abstract}
  Within a semiclassical description of above-threshold ionization
  (ATI) we identify the interplay between intracycle and intercycle
  interferences. The former is imprinted as a modulation envelope on
  the discrete multiphoton peaks formed by the latter. This allows to
  unravel the complex interference pattern observed for the full
  solution of the time-dependent Schr\"odinger equation (TDSE) in
  terms of diffraction at a grating in the time domain. These
  modulations can be clearly seen in the dependence of the ATI spectra
  on the laser wavelength. Shifts in energy modulation result from the effect of the long
  Coulomb tail of the atomic potential.
\end{abstract}

\maketitle

Tunneling ionization is a highly nonlinear quantum-mechanical
phenomenon induced by intense laser pulses ($\gtrsim
10^{14}\,$W/cm$^{2}$). Electrons are emitted by tunneling through the
potential barrier formed by the combination of the atomic potential
and the external strong field. Tunneling has recently attracted
increasing interest as a probe of the atomic and molecular structure
\cite{Meckel08,Gopal09,Arbo06prl}. Tunneling occurs within each optical cycle
predominantly around the maxima of the absolute value of the electric
field. The interference of the successive bursts of ejected electrons
reaching the same final momentum gives rise to features in
photoelectron energy and momentum distribution which are markedly
different from typical above-threshold ionization (ATI) spectra by
multi-cycle lasers.  This temporal double-slit interference has
recently been studied both experimentally \cite{Lindner05,Gopal09} and
theoretically \cite{Arbo06}. On the other hand, the ATI peaks
separated by a photon energy can be themselves viewed as an
interference pattern formed by electron bursts repeated each optical
cycle. Details of the interplay between these intra- and
intercycle interferences have not yet been clearly identified and
analyzed, to the best of our knowledge.

In this Rapid Communication, we study the influence of
different interference processes on ATI spectra generated by
multi-cycle laser pulses.  We clarify the underlying mechanism within
a simple one-dimensional model employing classical
trajectories. Within the framework of the strong-field approximation
(SFA) \cite{lewenstein} the qualitative features, the modulation of
the ATI peaks akin to the modulation of Bragg peaks by the structure
factor in crystal diffraction, can be unambiguously identified in the
ATI spectrum determined from the full solution of the
three-dimensional time-dependent Schr\"odinger equation (TDSE). The
multi-cycle laser pulse thus acts as a diffractive grating in the time
domain. Quantitative deviations between the SFA predictions and the
full TDSE can be traced to the Coulomb tail of the atomic potential
affecting this modulation. The latter opens up the opportunity to
observe effects of the atomic potential in easy-to-obtain
photoelectron spectra after ionization by multi-cycle laser pulses.

Our simple semiclassical model of photoelectron spectra is based on
the 1D \textquotedblleft simple man's model (SMM)\textquotedblright
\cite {Delone91,lewenstein,Chirila05}. Let us consider an atom
interacting with a linearly polarized laser pulse. The laser field
$F(t)$ is chosen to be of the form $F(t)=f(t)\sin \omega t$, with an
envelope function $f(t)$ corresponding to an $N$-cycle flat-top pulse
with a field strength of $F_{0}$ and with $m$-cycle linear ramp-on and
-off (we set $m=\frac{1}{2}$ in the following). The classical electron
trajectories $i$ ($i=1,...,2N$) for a final momentum $k$ are
characterized by their release times $t_{r}^{(i)}$ which satisfy,
\begin{equation}
k=-A(t_{r}^{(i)}),  \label{clpz}
\end{equation}
where $A(t)=-\int_{-\infty }^{t}dt^{\prime }F(t^{\prime })$ denotes
the vector potential divided by the speed of light, with
$A(t)=(F_{0}/\omega )\cos \omega t$ for the flat-top segment of the
pulse. In this study we focus on direct photoelectrons (without
rescattering) with energies $E \lesssim 2 U_p \,\,
(U_p=F^2_0/4\omega^2)$ which dominate the total ionization
probability. It should be noted that within the flat-top part of the
pulse and for a given value of $k$, the field strength upon ionization
$\left\vert F(t_{r}^{(i)})\right\vert $ is a constant independent of
the release time $t_r^i\;(i=1,...,2N)$.  Thus, assuming that the
ground-state depletion is negligible, the ionization rate
$\Gamma(t_{r}^{(i)})$ is identical for all the ionization bursts (or
trajectories) to first approximation and only a function of $\Gamma
(k)$. Consequently, the momentum distribution $P(k)$ can be written
as
\begin{equation}
P(k)=\Gamma(k)\left\vert \sum_{i=1}^{2N}e^{iS(t_{r}^{(i)})}\right\vert ^{2},
\label{interf}
\end{equation}%
where $S$ denotes the Volkov action \cite{Volkov} 
\begin{equation}
S(t)=-\int_{t}^{\infty }dt^{\prime }\left[ \frac{(k+A(t^{\prime }))^{2}}{2}%
+I_{p}\right] \;,  \label{action}
\end{equation}%
with $I_{p}$ being the ionization potential. The key to the analysis
of intracycle and intercycle interferences is that the sum over
interfering trajectories (Eq.\ \ref{interf}) can now be decomposed into
those associated with two release times within the same cycle and
those associated with release times in different cycles (see Fig.~1)
\begin{equation}
\label{eq:four}
\sum^{2N}_{i=1} e^{iS(t_r^i)}=\sum^{N}_{j=1} \, 
\sum^{2}_{\alpha=1} e^{iS (t_r^{(j, \alpha)} ) } \, .
\end{equation}
Accordingly, Eq. (\ref{interf}) becomes
\begin{equation}
P(k)=\Gamma (k) \cos ^{2}\left( \frac{\Delta S}{2}\right) \left\vert
\sum_{j=1}^{N}e^{i\bar{S}_{j}}\right\vert ^{2} =\Gamma (k) F (k) B (k) 
\label{intra+inter}
\end{equation}%
where $\bar{S}_{j}=[S(t_{r}^{(j,1)})+S(t_{r}^{(j,2)})]/2$ is the
average action within each cycle and $\Delta
S=S(t_{r}^{(j,1)})-S(t_{r}^{(j,2)})$ is the difference. Note that
$\Delta S$ is independent of $j$. Eq.\ (\ref{intra+inter}) is
structurally equivalent to the intensity for crystal
diffraction: the factor $F(k) = \cos ^{2}\left( \Delta S /2\right)$ represents the form factor (on structure) accounting for interference modulations due to the internal
structure within the unit cell while the second factor $B(k)=\left\vert \sum_{j=1}^{N}e^{i\bar{S}_{j}}\right\vert ^{2}$ gives
rise to Bragg peaks due to the periodicity of the
crystals. Alternatively, Eq.\ (\ref{clpz}) may be viewed as a
diffraction grating in the time domain consisting of $N$ slits and
with $F(k)$ being the diffraction factor for each slit. Interferences
between different slits correspond to intercycle interferences while
$F (k)$ represent intracycle interferences.

\begin{figure}[tbp]
\centerline{\includegraphics[width=8.3cm]{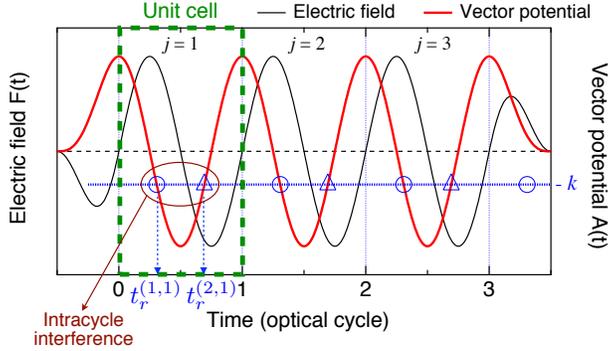}} 
  \caption{(color online) Electric field $F(t)$ (left axis) and
    vector potential $A(t)$ (right axis) of a flat-top
    pulse ($N=4,\, m=\frac{1}{2}$). The electron emission times for a
    given final momentum $k$ are marked by circles ($t_r^{(j,1)}$) and
    triangles ($t_r^{(j,2)}$). Each circle-triangle pair determines
    the structure factor $F(k)$ and leads to intracycle interference
    while the periodic train of such pairs gives rise to intercycle
    interference. Each optical cycle can be viewed as
    ``unit cell'' of the time
  lattice.}
\label{fig:field}
\end{figure}

\begin{figure}[tbp]
\includegraphics[width=8.3cm]{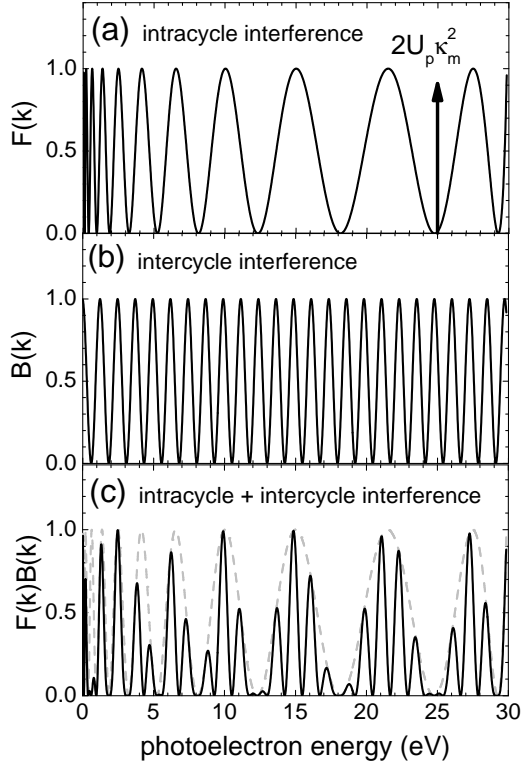}
\caption{Semiclassical intracycle and intercycle interference pattern
  (color online). (a) Intracycle interference pattern given by the
  structure factor $F(k)$.  (b) Intercycle interferences with
  ``Bragg'' peaks given by $B(k)$ in Eq. (\protect\ref{eq:intercycle}). (c)
  Total semiclassical interference (Eq. (\protect\ref{intra+inter}%
  )) for $N=2$. The laser wavelength and intensity are $1000$ nm and
  $1.6\times 10^{14}\,\mathrm{W/cm}^2$, respectively.  To stress the
  interferences, we set $\Gamma(k)=1$ and normalize the respective
  maxima of $F(k)$ and $B(k)$ to unity, see Eq.(\ref{intra+inter}). }
\label{fig:intra+inter}
\end{figure}

We show in the following that the interplay between $B (k)$ and $F
(k)$ controls the direct ATI spectrum (Fig.~2).  Note that the peak of
the intracyle structure factor $F(k)$ are not equispaced in energy
(shown in Fig.~2~(a) for hydrogen with $I_P=0.5\, \mathrm{a.u.}$).  The
separation of consecutive peaks is larger at intermediate energies
than near the classical boundaries $E=0$ and $E=2U_p$ ($U_p=0.5\,
\mathrm{a.u.}$ in the present case). It can be analytically shown that
the separation between adjacent peaks $\Omega (k)$ as a function of the
final momentum is given by
\begin{equation}
\Omega (\kappa )=\frac{\omega \pi \kappa \sqrt{1-\kappa ^{2}}}{1-\kappa
^{2}+\gamma ^{2}/2-\kappa \sqrt{1-\kappa ^{2}}\arccos \kappa },
\label{T(Kappa)}
\end{equation}%
where $\gamma =\sqrt{2I_{p}}\omega /F_{0}$ is the Keldysh parameter,
and $%
\kappa =\omega k/F_{0}$ is the scaled dimensionless momentum. $\Omega
(\kappa )$ reaches a maximum at $\kappa _{m}=\sqrt{1+\frac{\gamma
    ^{2}}{2}-%
  \frac{\gamma }{2}\sqrt{2+\gamma ^{2}}}$, the position of which is
indicated by a vertical arrow in Fig.~\ref{fig:intra+inter}~(a).  Such
intracycle interference patterns have been recently experimentally
observed for ultrashort near-single cycle IR pulses
\cite{Gopal09,Lindner05} and theoretically analyzed \cite{Arbo06}.
The intercycle interferences pattern (Fig.\ 2b) is, by contrast,
equispaced and corresponds to the well-known ATI pattern with peak spacing
corresponding to one-photon transitions $(\hbar \omega)$. For the case of $N=2$ in Fig. 2(b), $B(k)$ is given by
\begin{equation}
  \label{eq:intercycle} 
B(k)=4\cos^{2}\left( \frac{S\left( t_{r}^{(1)}\right) -S\left( t_{r}^{(1)}+2\pi
/\omega \right) }{2}\right) \;.
\end{equation}

\begin{figure}[tbp]
\includegraphics[width=10.3cm]{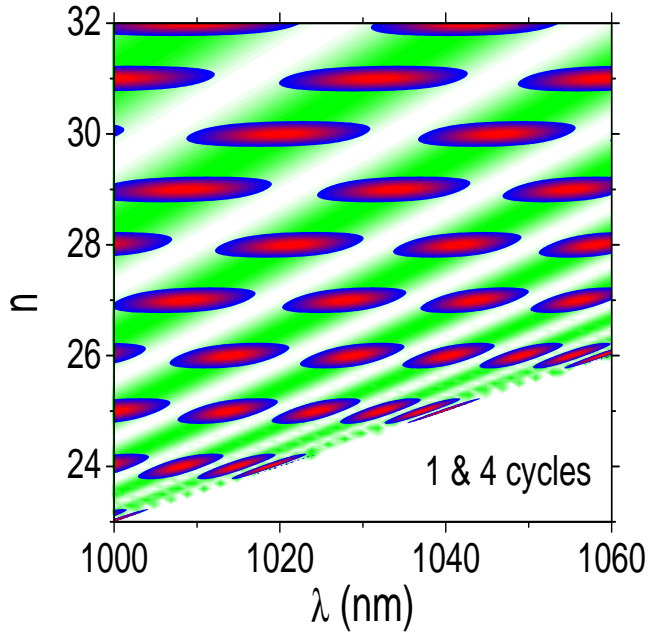}
\caption{(color online) Semiclassical total ionization probability as
a function of the laser wavelength $\lambda$ and the ATI order
$n$ (see text). Shown is the spectrogram for both a one-cycle pulse
(green stripes) and and a four-cycle pulse (red
islands). $F_{0}=0.0675\,\mathrm{a.u.} \; (1.6\times
10^{14}\,\mathrm{W/cm}^2)$.}
\label{fig:2D_SC}
\end{figure}

In multi-cycle photoelectron spectra with $N\geq 2$, both intracycle
and intercycle interferences are simultaneously present. In
Fig. \ref{fig:intra+inter}~(c), the resulting energy distribution
calculated from the time grating (Eq. (\ref{intra+inter})) is shown
for $N=2 $. The multiphoton peaks are modulated by the intracycle
interference structure factor. The ATI peaks become narrower as the
number $N$ of optical cycles increases approaching, in this way, $\delta$-peaks for
infinitely long pulses. On the contrary, the \textit{intracycle}
modulation is independent of the number of cycles. Thus, the sub-cycle
interference, previously studied with near-single-cycle ultrashort
pulses, is embedded and visible in ATI spectra for multi-cycle pulses,
a feature apparently up to now not fully recognized. This effect
becomes more transparent when we study the parametric variation of the
photoelectron spectrum. In Fig.~\ref{fig:2D_SC} we show the
spectrum 
expressed in units of the (new) photon number $n=(E+I_p+U_p)/\omega$,
as a function of laser wavelength $\lambda$.  For $N=4$
cycles the horizontal stripes in this two-dimensional interferogram
peaking near integer values of $n$ represent ATI peaks due to
intercycle interference. Superimposed are tilted stripes controlled by
the intracycle interference, which are also visible for the $N=1$
cycle pulse. We note that a very similar two-dimensional interferogram
has recently been found in atom-surface diffraction \cite{Aigner}.

In view of the fact that the present intracycle and intercycle
interference structures are derived from a simple one-dimensional SFA
model neglecting realistic features of the atomic potential, we test
its predictions against a full numerical 3D TDSE solution
\cite{ar:tong97} for hydrogen. The resulting interferogram for
photoelectrons (Fig.\ 4) agrees qualitatively remarkably well with the
one-dimensional SFA model (Fig.\ 3). The intracycle modulation is best
seen when the angular acceptance of the photoelectron spectrum is
restricted to a cone of small angles around the polarization axis
($\theta=10^\circ$ in Fig.\ 4a) while it is somewhat blurred but still
visible in the total spectrum (Fig.\ 4b). We found similar
interference patterns for rare gases such as argon, though the details
depend on the atomic potential.
Projecting the two-dimensional distribution onto the $\protect\lambda$
axis (Fig.\ 4c) results in a regular modulation pattern on a fine
$\protect\lambda$ scale which can be traced back to the combined
effect of inter- and intracycle interferences. This oscillations
closely resemble those previously observed for the $\protect\lambda$
dependence of HHG \cite{Schiessl07,Schiessl08,Ishikawa09}.

\begin{figure}[tbp]
\centerline{\includegraphics[width=8.3cm]{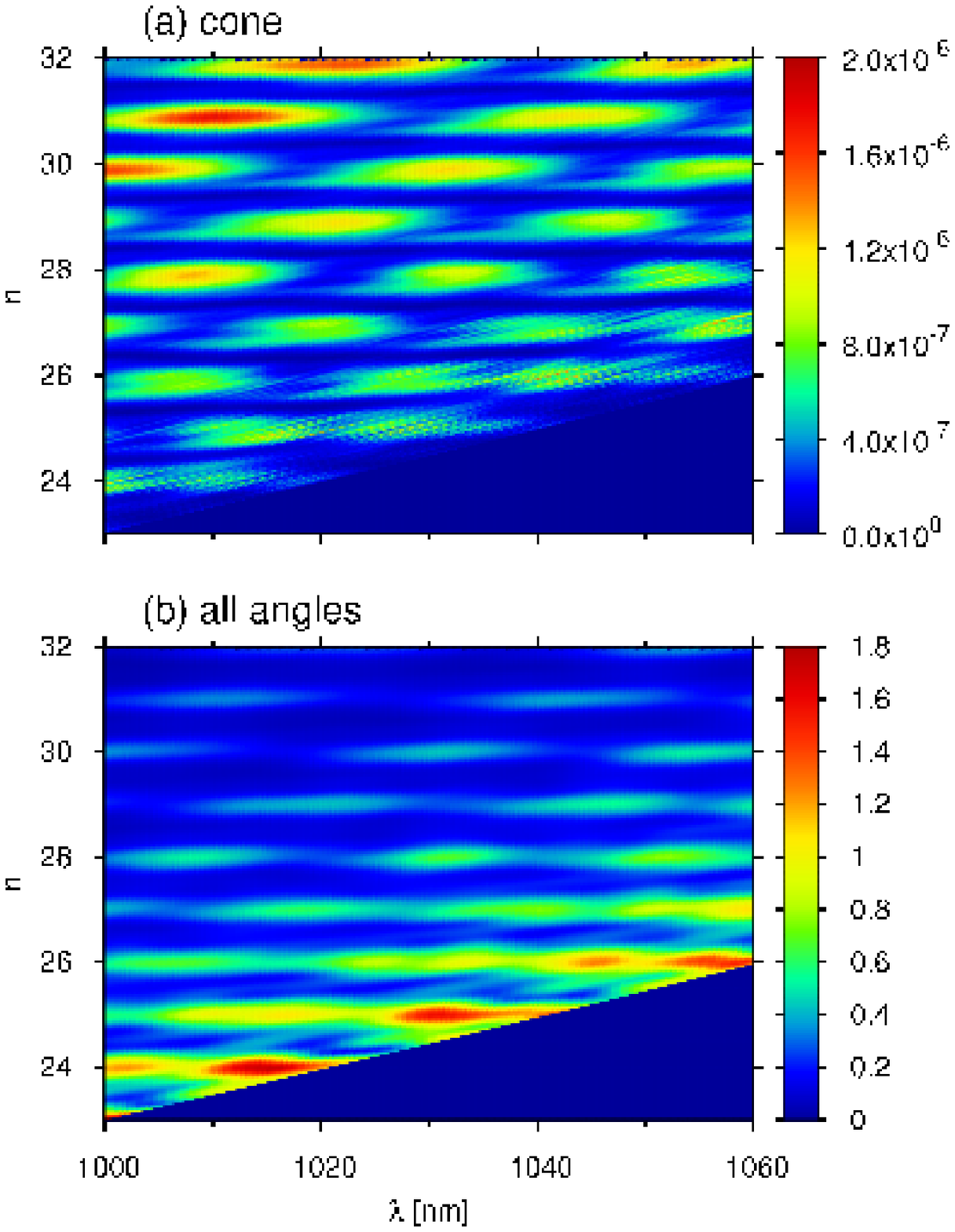}} 
\centerline{\includegraphics[width=8.3cm]{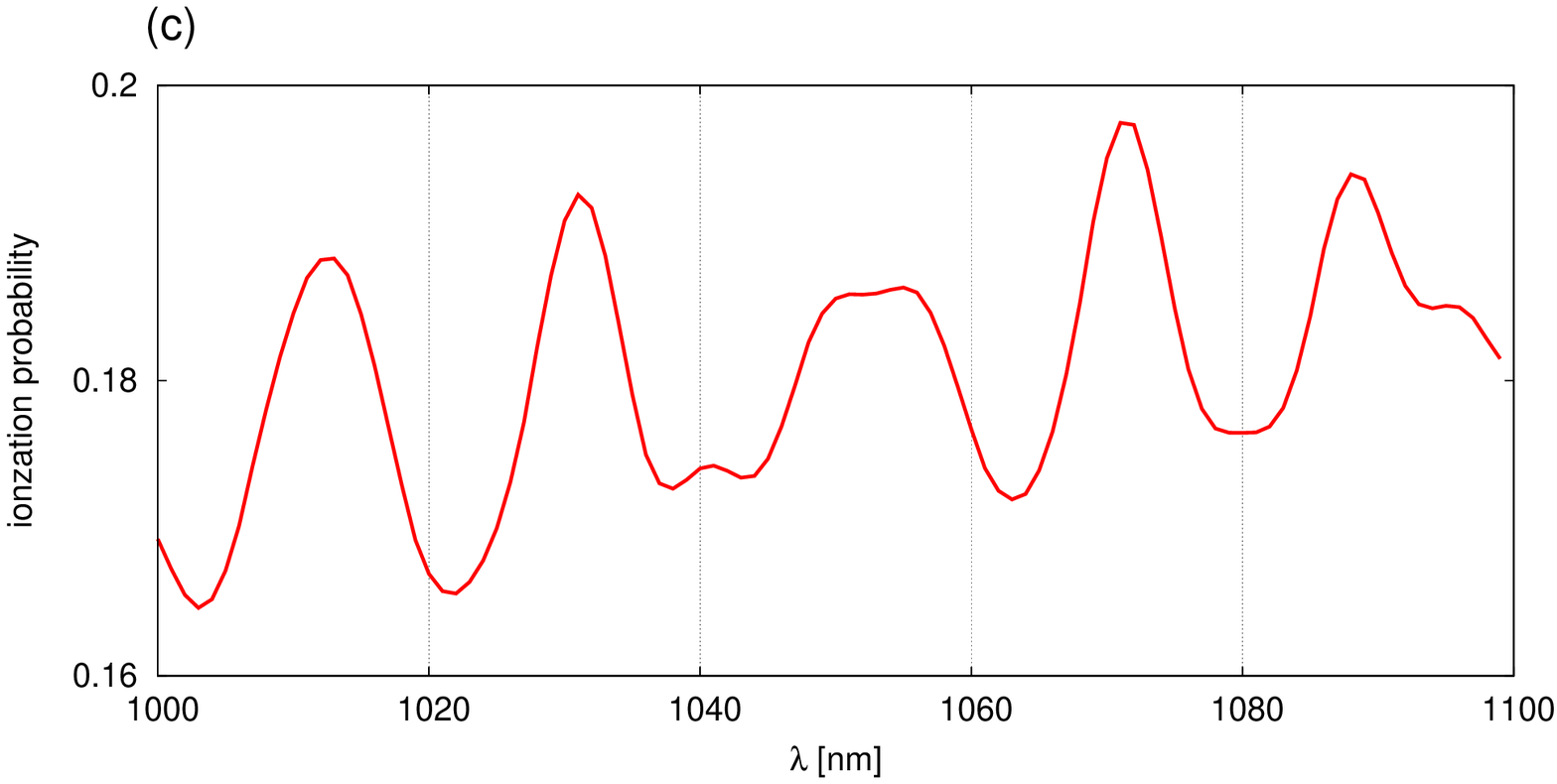}}
  \caption{(color online) Two-dimensional interferogram as a function
    of laser wavelength $\protect\lambda$ and ATI order $n$ (see text)
    calculated by using the TDSE for hydrogen for a four-cycle pulse with
    $F_{0}=0.0675\,\mathrm{a.u.} \ (1.6\times
    10^{14}\,\mathrm{W/cm}^2)$. (a) Emission into a cone of $10^\circ$
    around the polarization axis and (b) for all angles. (c)
    Energy-integrated total ionization yield as a function of
    $\protect\lambda$.}
\label{fig:2D_TDSE}
\end{figure}

\begin{figure}[tbp]
\centerline{\includegraphics[width=8.3cm]{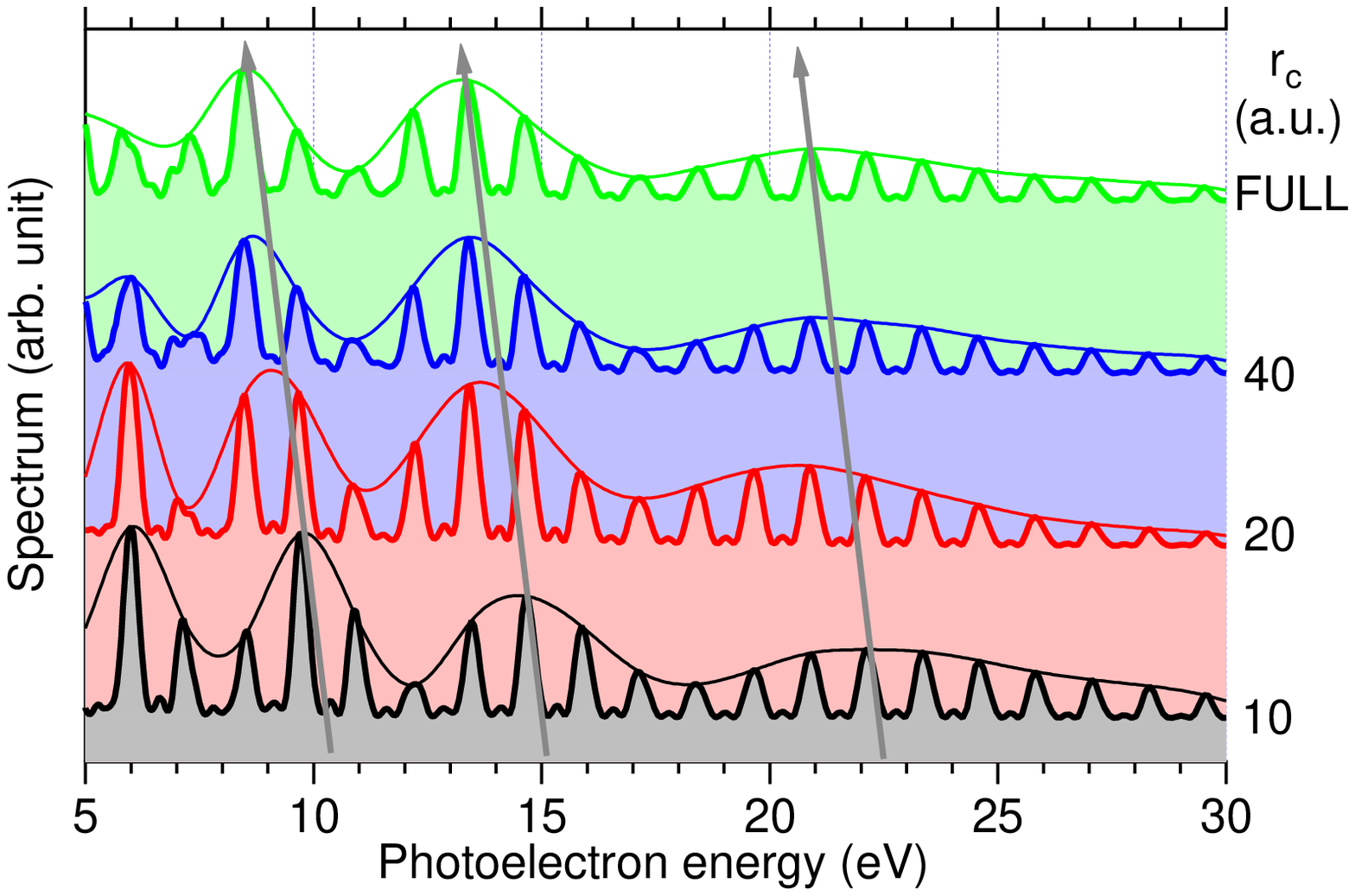}}
\caption{(color online) Comparison of the photoelectron spectra within
  the $10^\circ$ cone, calculated with the full Coulomb potential
  (marked as \textquotedblleft full\textquotedblright) with the
  truncated Coulomb potentials for varying values of $r_c$ as
  indicated. The laser wavelength is 1000 nm. Other laser parameters
  are the same as in Fig.\ \protect\ref%
  {fig:2D_TDSE}. Thin solid lines are approximate peak envelopes.}
\label{fig:TDSE_tr}
\end{figure} 

Experimentally, more easily accessible is the variation of the
photoelectron spectra as a function of the energy at fixed wavelength,
i.e., projection of the 2D interferogram onto the $\lambda$ axis (Fig.\
5). The photoelectron spectrum clearly reflects the modulation of the
regularly spaced ATI peaks (intercycle interference) by intracycle
interference, in complete agreement with the prediction of a time
grating (Eq. (\ref{intra+inter}). It is now highly instructive to probe
this interference pattern for its dependence on the Coulomb potential
neglected in the SFA derivation of the time grating. Within the full
TDSE we can explore Coulomb-tail effects by using a screened Coulomb
potential,
\begin{equation}  \label{eq:truncated_potential}
V(r) = \left\{%
\begin{array}{ll}
-\frac{1}{r} & (r < r_c) \\ 
-\frac{e^{-(r-r_c)/r_d}}{r} & (r > r_c)%
\end{array}%
\right.\qquad ,
\end{equation}
where the parameter $r_d$ characterizes the width of the transition
from the full to the screened Coulomb potential. We use $r_d=10~$a.u.,
and $r_c$ is varied between 10 and 70~a.u. For these parameter values,
the ionization potential and the first excitation energy remain
unchanged to an accuracy ranging between $10^{-3}$ and $10^{-9}$.
Consequently, the position of the ATI peaks remains unchanged as a
function of $r_c$. However, the modulation envelope is systematically
and almost rigidly shifted towards lower energies under the influence
of the long-range Coulomb tail. The Coulomb phase does not affect the
Bragg peaks but changes the structure factor $F (k)$ of the time
grating in Eq. (\ref{intra+inter}).
Thus, the Coulomb potential is responsible for a shift in the
positions of the intracycle interference stripes, visible in
Figs. \ref{fig:2D_SC} and \ref{fig:2D_TDSE}(a).
It should be emphasized that the shift is equally
significant even for higher photoelectron energies, for which one
might expect the SFA to be better justified.  We note parenthetically
that such shifts can be partially accounted for by including the
Coulomb-Volkov phase into the semiclassical theory
\cite{Ishikawa09,Arbo}.

In summary, we have identified the interplay between the intra- and
intercycle interferences of electron trajectories in ATI spectra by
intense multicycle laser pulses.  The former, which carries
information on attosecond subcycle dynamics of the electron cloud
\cite{Gopal09}, is imprinted as a pulse-length independent modulation
of multiphoton peaks formed by the latter. This modulation is even
more clearly visible in the wavelength dependence of ATI spectra,
which could be measured with state-of-the-art tunable sources based,
e.g., on optical parametric chirped pulse amplification. While the
general features of the interplay between the intracycle and
intercycle interferences is well explained by the simple semiclassical
theory, the intracycle modulation envelope is shifted by the Coulomb
tail which emphasizes the effect from the atomic potential.

This work was supported by CONICET and PICT2006-00772 of ANPCyT (Argentina),
SFB 016 ADLIS and P15025-N08 of FWF (Austria), and by the EU project
HPRI-2001-50036. K.L.I. gratefully acknowledges financial support by the
JST-PREST program, the MEXT Grant No. 19686006 (Japan), 
and the Advanced Photon Science Alliance (APSA) project (Japan) and KS
acknowledges support by the IMPRS program of the MPQ (Germany)

\end{document}